\documentstyle[12pt,aasms4]{article}

\def\fmag{\hbox{$.\!\!^{\rm m}$}}
\def\fdeg{\hbox{$.\!\!^\circ$}}

\slugcomment{}

\lefthead{Beckwith et al.}
\righthead{An Emission Line Galaxy at $z = 2.43$}

\begin{document}

\title{An Infrared Emission Line Galaxy at {\boldmath $z = 2.43$}
       \footnote{Based partially on observations obtained at the W. M. Keck
       Observatory, which is operated jointly by the California Institute of
       Technology and the University of California.}}

\author{S. V. W. Beckwith}
\affil{Max-Planck-Institut f\"ur Astronomie, Heidelberg, Germany\\
Electronic mail: svwb@mpia-hd.mpg.de}

\author{D. Thompson}
\affil{Max-Planck-Institut f\"ur Astronomie, Heidelberg, Germany\\
Electronic mail: djt@mpia-hd.mpg.de}

\author{F. Mannucci}
\affil{C.A.I.S.M.I.--C.N.R., Florence, Italy\\
Electronic mail: filippo@arcetri.astro.it}

\and

\author{S. G. Djorgovski}
\affil{California Institute of Technology, Pasadena, California, USA\\
Electronic mail: george@oracle.caltech.edu}

\begin{abstract}

An object discovered during an infrared survey of the
field near the quasar B2 0149$+$33 has an emission line at
2.25\,$\mu$m that we interpret as H$\alpha$ at a redshift of 2.43.
The K-band image shows two compact components 10\,kpc apart
surrounded by more extended emission over $\sim 20$\,kpc.  The H$\alpha$
emission appears to be extended over $\sim 15$\,kpc (2$^{\prime\prime}$)
in a coarsely sampled (0\farcs8/pixel) image.  The star formation rate
may be as high as 250 -- 1000\,M$_\odot$\,yr$^{-1}$, depending on the
extinction.  Alternatively, the line may be powered by an active nucleus,
although the probability of serendipitously discovering an AGN in the survey
volume is only $\sim 0.02$.  The increasing number
of similar objects reported in the literature indicate that they
may be an important, unstudied population in the high redshift universe.

\end{abstract}


\section{Introduction}

Discovering the properties of galaxies at redshifts greater than one
requires techniques that can readily distinguish the high redshift
objects from those at lower redshift that predominate in any deep
image.  Multi-wavelength approaches are especially fruitful and have
been dominated to date by optical and radio surveys.  The use of
photometric redshifts based on the strong Lyman break redshifted into
the optical band has been one of the most successful of these
techniques (Steidel et al. 1996, and references therein).

Methods using infrared images are beginning to uncover objects not
easily discovered with optical or radio methods that may,
nevertheless, constitute a significant fraction of the high redshift
population. These include objects distinguished by unusually red
colors (Elston, Rieke, \& Rieke 1988; Soifer et al. 1994; Graham et
al. 1994; Hu \& Ridgway 1994; Cowie et al. 1994; Dey, Spinrad, \&
Dickinson 1995) and objects with emission lines redshifted to infrared
wavelengths (Songaila et al. 1994; Thompson, Mannucci, \& Beckwith
1996, hereafter TMB96; Malkan, Teplitz, \& McLean 1995; Bechtold et
al. 1997). These objects have been interpreted as elliptical galaxies
(Graham et al. 1994; Hu \& Ridgway 1994; Dunlop et al. 1996), young
``protogalaxies'' undergoing bursts of star formation (Eisenhardt \&
Dickinson 1992; Graham \& Dey 1996; Malkan, Teplitz, \& McLean 1996;
Yee et al. 1996; Bechtold et al. 1997), or active galactic nuclei
(Cowie et al. 1994; Dey, Spinrad, \& Dickinson 1995).  All of these
populations could be significant for cosmology, since the first case
implies massive galaxy formation at redshifts greater than $\sim$3,
the second indicates a substantial population of young galaxies that
can be discovered only in infrared surveys, and the third implies a
population of infrared bright active galactic nuclei (AGN) comparable
in number density to the populations of AGN discovered by more
traditional methods.  Only a few such objects have been studied in
enough detail to reveal redshifts, source morphologies, and colors.

In the course of our survey for emission line galaxies (TMB96), an
object was discovered near the quasar B2 0149$+$33 that had an
emission line at 2.25\,$\mu$m, the same wavelength as the quasar's
H$\alpha$ line at a redshift of 2.43; we call this object TMB 0149 -
cK39, or simply cK39. A spectrum between 1.5 and 2.4 $\mu$m, presented 
and discussed here, confirmed the presence of an emission line nearly 
coincident in wavelength with that of the quasar. The object is very 
red, making it unusual among known, distant galaxies.  This paper 
describes the results of these observations and suggests that such 
objects may be common but previously unobservable owing to the lack 
of optical emission.

\section{Observations}

The object was discovered in the course of a survey that covered 250
square minutes of arc at wavelengths where the H$\alpha$ line would
appear at redshifts between about 2.1 and 2.7 (TMB96).\footnote{the
survey covered a total of 276 square arcminutes including lines other
than H$\alpha$}  The total co-moving volume was 3$\times10^4
h_{50}^{-2}$ Mpc$^3$ ($h_{50}$ $=$ H$_0$ / 50 km s$^{-1}$ Mpc$^{-1}$,
q$_0$ $=$ 0.5 are assumed throughout this paper).  Within this volume,
there was only one candidate discovered by comparing images in broad
and narrow band filters, searching for emission line objects that were
brighter in the narrow band image and more than 3$\sigma$ above the
background.  

Simultaneous spectra in the H and K bands of the candidate and the 
quasar, B2 0149$+$33, were taken with the grating spectrometer, CGS4, at
the UKIRT telescope on the nights of UT 1996 January 12 to 14.  The 
weather was not photometric, with variable seeing at $\sim$2\farcs5, 
measured from the quasar continuum.  The objects were dithered along 
the slit by 10\arcsec\ to 20\arcsec\ and the spectra at these two 
positions subtracted to eliminate sky emission.  The 2-pixel wide 
slit (2\farcs4)produced a spectral resolution of 195$\lambda$\,($\mu$m) 
or 438 at the position of the quasar's H$\alpha$ emission line.  The 
spectra of the quasar and cK39 were extracted using variance weighting 
in the IRAF apextract package, summing across a 2-pixel wide window 
along the slit.  The poor and variable seeing, combined with uncertainties 
in the positioning of the slit, preclude an accurate measurement of the 
line flux from this spectrum.

Figure~\ref{fig1} shows the resulting spectra, centered on the H$\alpha$ 
line of the quasar.  The quasar spectrum shows a strong H$\alpha$ line
redshifted to 2.25\,$\mu$m, while a faint emission line at the same 
wavelength is apparent at the positions of cK39.  The bottom part of the 
figure shows the extracted spectrum of the line in cK39 at full resolution, 
overlaid with a scaled plot of the quasar's H$\alpha$ line.  The 
formal signal-to-noise ratio of the emission line is 6.3, inadequate to 
conclude more than that an emission line is present, and it is impossible 
to separate the line into individual components, from H$\alpha$ and [NII], 
for example.  Although the emission line in cK39 appears somewhat broad, 
with a formal line width of 2200\,km\,s$^{-1}$ (FWHM), it may be unresolved
within the considerable uncertainties.  The line could thus arise from a 
Galactic object with essentially zero width or an active nucleus with broad 
lines. 

\placefigure{fig1}	

Figure~\ref{fig2} shows images taken on the nights of UT 1997 September 
9 (K band) and UT 1997 October 3 (R band) with the 10\,m W. M. Keck
telescopes, obtained in order to determine the morphology of cK39 more 
accurately.  The K band image shows that cK39 is a pair of extended
sources 1\farcs 3 apart, cK39E \& cK39W.  The western
component, cK39W, has K $=$ 20.27 $\pm$ 0.06 in a 0\farcs 5
radius aperture; the eastern component is lower surface brightness with
K $=$ 20.57 $\pm$ 0.07 in the same aperture.   Each component is more
extended than the 0\farcs 6 seeing profile: cK39W is
0\farcs 75 and cK39E is 0\farcs 9 (FWHM), uncorrected for
the seeing.  The continuum-corrected line emission did appear to be
extended in the original survey images, although it is difficult to make 
an unambiguous determination of the extent in the coarsely sampled survey
data.  We were, however, unable to reproduce the distribution of light
in cK39 using one or two point sources scaled down from a much
brighter star in the same image.  The extended morphology suggests that 
the object is a pair of galaxies perhaps caught in the process of merging.  
The candidate, cK39, is barely detected in the R band image 
(0\farcs 6 seeing) as a low surface-brightness diffuse object.  
It is clearly extended along an east-west line, but the two components 
cannot be readily distinguished in the image for the purposes of photometry.  
The center of light is close to cK39W, however, indicating that the western 
component is brighter and bluer than cK39E.  

\placefigure{fig2}	

Table~\ref{table1} lists the parameters of the various observations along 
with the results, including those from TMB96, where relevant.  The 
magnitudes of the individual components were derived using 
1\arcsec\ apertures, and they do not include extended light that 
contributes to the total magnitude.  The line fluxes were derived from a 
comparison of the original narrow and broad band images with 
0\farcs 8 pixel resolution.  The systematic uncertainties of 
deriving photometric magnitudes between the narrow and broad band filters 
combined with the problem that the center of the narrow band filter is at 
one edge of the broad band filter requiring uncertain color corrections 
means that the tabulated line flux may be 50\% brighter or fainter 
(1$\sigma$) than the listed value.

\placetable{table1}	

\section{Discussion}

The coincidence of line emission between the quasar and cK39 and the
$\sim$~2$^{\prime\prime}$ size of cK39 suggests that the line is
H$\alpha$+[NII] and cK39 is at a redshift of 2.43.  If the object were in
the Galaxy or at very low redshift, the only plausible identification
would be the v=2$-$1 S(1) line of H$_2$ at 2.248\,$\mu$m, a weak line
always accompanied by a series of other lines (e.g. Shull \& Beckwith
1982) that are not seen in our spectrum.  The other strong lines from
an extragalactic object that are likely to appear at these wavelengths
are Br$\gamma$ ($\lambda_0 = 2.1656$\,$\mu$m), P$\alpha$ ($\lambda_0 = 
1.8751$\,$\mu$m), and [Fe II] ($\lambda_0 = 1.6435$\,$\mu$m)\footnote{Other 
possible lines exist, but they are invariably weaker than the hydrogen 
lines}. If the line were Br$\gamma$ at $z=0.039$, the P$\alpha$ line would 
be seen at 1.948\,$\mu$m, within the range of our spectrum and strong 
enough to show up even through the poor telluric transmission.  The 
shock-excited line [Fe II] (at $z=0.369$) can be strong in merging systems 
(Elston \& Maloney 1990; Lester, Harvey, \& Carr 1988), as suggested by 
the morphology of cK39 in the K band, though Pa$\beta$ would be visible 
in our spectrum at 1.755\,$\mu$m.  P$\alpha$ is thus the only reasonable 
alternative to H$\alpha$+[NII] to explain the spectrum of cK39.  For 
the P$\alpha$ interpretation, however, one would also have to consider 
that the cK39 system would then be very low luminosity, several 
magnitudes fainter than an L$^*$ galaxy.  

Table~\ref{table2} gives some of the properties of the object assuming 
the line is either P$\alpha$ or H$\alpha$.  The star formation rates 
have been estimated using Kennicutt's (1983) empirical relationship 
between H$\alpha$ line luminosity and star formation rate (SFR), assuming 
zero extinction, no contribution from an active nucleus.  We ignore the 
possible contribution of [NII] to the H$\alpha$ line flux following 
the arguments of Bechtold et al. (1997) which also apply to cK39 for 
the purposes of this calculation. Table~\ref{table2} includes estimates 
of the visual extinction, A$_{\rm V}$, local to cK39 needed to produce the 
observed R$-$K$^\prime =$ 5.5 from various template galaxy spectra (Bruzual 
\& Charlot 1993; Coleman, Wu, \& Weedman 1980).   Galactic extinction 
at the lattitude of cK39, -27\fdeg3, is estimated to be 0\fmag3 at 
R and 0\fmag1 at K (or K$^\prime$), which is negligible considering 
other uncertainties.  
An extinction of A$_{\rm V}$ $\sim 1.7$ (mid-range for different galaxy 
models) is required to redden a young star burst galaxy to produce the 
observed color at a redshift of 2.4, while an extinction A$_{\rm V} >  
4^{\rm m}$ would be required to produce the observed colors if cK39 is 
at the lower redshift.

\placetable{table2}	

Table~\ref{table2} also lists the minimum and maximum volume densities of 
objects within the $\pm$2$\sigma$ probability band at the different redshifts,
computed by setting the Poisson probability to 0.046 (2$\sigma$).  If
the line is H$\alpha$+[NII], the probability might be substantially 
enhanced if objects exhibit even mild clustering with quasars.  The 
maximum separation between cK39 and quasar is 9\,Mpc, adopting the
1$\sigma$ maximum redshift difference of 0.01 allowed by the spectra 
($\Delta\lambda = 0.002\pm 0.0045\,\mu$m for $\lambda_0 = 0.656\,\mu$m); 
the minimum projected separation is 390\,kpc.  In this case, the average
density of such objects might be lower by a factor of ten or more than
$\rho_{\rm min}$ without affecting the probability estimates. Our
survey covered a much larger volume near redshift 2.4 than at
redshifts below 1, and the global star formation rate appears to be
larger at $z\sim$2.4 than at $z\sim$0.2 (Madau et al. 1996).  The chance
of discovering any kind of exotic object should thus be greater at higher
redshifts.  We therefore believe that this line is H$\alpha$+[NII].

The object cK39 shares some properties with other infrared-bright
galaxies at high redshift, MS 1512$-$cB58 (Yee et al. 1996; Bechtold
et al. 1997), MTM 095355$+$545428 (Malkan, Teplitz, \& McLean 1995,
1996), Hawaii 167 (Cowie et al. 1994, now known to be a quasar), 
and HR 10 (Graham \& Dey
1996).  Characteristics of these objects are compared in Table~\ref{table3}.
All have large apparent star formation rates as compared
to the 5 - 100\,M$_\odot$\,yr$^{-1}$ typical of young galaxies in the
redshift range 1 - 3 found in ''dropout'' samples (Steidel et al. 1996; 
Pettini et al. 1997; Dickinson 1998).
If cK39 is a galaxy or ongoing merger-induced starburst,
powered solely by photoionization from massive stars, it has one of
the highest star formation rates seen in high redshift objects.   

\placetable{table3}	

Graham \& Dey (1996) had difficulty reconciling properties of HR10
that are nearly identical to cK39.  HR10 was discovered because of its
color and only later discovered to have a strong emission line; the
opposite is the case for cK39. But the emission line made it possible
to argue that HR10 could not be an elliptical galaxy as proposed by Hu
\& Ridgway (1994) on the basis of the colors alone, and must contain
young stars or an active nucleus.  The irregular shape and the strong
rest-frame blue (observed 1\,$\mu$m) flux density make HR 10 more
likely to be a reddened Sb galaxy with A$_{\rm V} \sim$ 1.8 $-$ 2.5.
Although cK39 was discovered first via its emission line, it is also
a red object with similar properties and contradictions as HR10.

Correcting the 240 $h_{50}^{-2}$ SFR of cK39 for the extinction
derived from the red color yields a total SFR which could be as
high as 10$^3$\,M$_\odot$\,yr$^{-1}$.  While the 10\,kpc separation
of the two components is consistent with a relative velocity of
100\,km\,s$^{-1}$ over 0.1\,Gyr, starbursts of this magnitude
cannot last for long before exhausting their material - an entire
elliptical galaxy could be created in 0.1 Gyr at the
10$^3$\,M$_\odot$\,yr$^{-1}$ rates implied by the observations.
In addition, the age of the universe at $z=2.43$ is about 3 Gyr for
our adopted cosmology.  If the lifetime of the object as observed is
only 0.1 Gyr, there would have to be roughly 30 times as many of these
objects as estimated in Table~\ref{table2} to make the probabilities 
consistent.  The density of the parent population would then be $5 \times
10^{-4}$\,Mpc$^{-3}$, about equal to the number of galaxies brighter
than L$^*$ using the usual Schechter function (Schechter 1976).  This
density is quite high and would imply that objects like cK39 could
account for the bulk of the star formation in the early universe.
Many galaxies would create their stars in a few, short but intense bursts.
However, there are currently no well-documented starbursts of this
magnitude which have not later been shown to contain an active nucleus,
and a number of surveys targetting intense starbursts at high redshift
have failed to turn up any significant population of objects.

A viable alternative to a massive starburst, and one which is
consistent with the data, is that an AGN produces some or all of the
H$\alpha$+[NII] emission.  The strength of
the H$\alpha$+[NII] line would imply a medium luminosity AGN that is quite
commonly found at redshifts $\sim$2, and would reduce or eliminate the
need to invoke such large star formation rates.  While the line emission
in cK39 appears extended in the narrowband image, there are a number of
high redshift radio galaxies with significantly extended line emission
powered by the active nucleus or radio jets without large amounts of
star formation.  The line width is also consistent with AGN line widths,
as well, and active nuclei are often found in disturbed systems
consistent with an ongoing or recent merger.  It is 
unlikely to find an AGN within the survey volume, however; the a priori
probability is of order 0.02, assuming no extinction at K, a spectral 
index of -0.5 to give an absolute magnitude M$_{\rm B} = -26.3$, and
Bolye's (1991) luminosity function for quasars.  
The original survey (TMB96) specifically targeted the environments
around known quasars and radio galaxies, so this probability
estimate ignores any possible contributions from clustering.  In the
Keck K-band image (Figure~1), both components appear extended,
and neither appears strongly dominated by the presence of a point source.

A third possibility is that a faint foreground galaxy or cluster
amplifies the line flux from cK39 through gravitational lensing.  The
luminous object, FSC 10214$+$4724 (Rowan-Robinson et al. 1991), is an
example of how gravitational lensing can distort the interpretation of
data taken with low spatial resolution and poor signal-to-noise ratios
(Goodrich et al. 1996).  To magnify by a factor of order 10, the
foreground galaxy must be almost coincident with cK39 along the line
of sight.  The faint R band detection could be such a galaxy and might 
then redden the light from cK39.  If so, it would be fainter than
typical lensing galaxies found to date.  The Keck images show no hint
of the arc morphology that describes most strongly lensed objects nor
an obvious foreground lens, even though the seeing-limited spatial
resolution is quite good, less than 0\farcs7.  We note that 
the apparently symmetric galaxy, MS 1512$-$cB58, exhibits arcs 
in an HST image and is, therefore, a gravitational lens (Bechtold 1997, 
personal communication).

\section{Conclusions}

The emission line object, TMB 0149 $-$ cK39, appears to be a pair of
galaxies undergoing a merger at redshift of 2.4.  If cK39 derives a
substantial part of its luminosity from star formation, the formation
rate is as high as 1000 M$_\odot$ yr$^{-1}$, an exceptionally large
value that is rarely seen in other starforming galaxies.
On the other hand, the system could contain at least one active nucleus,
perhaps with some contribution from star formation.  The individual
components appear to be 6 and 7 kpc in extent with the centers separated
by 10 kpc, consistent with a merger-induced fuelling of the nuclear
activity.  An extinction of A$_{\rm V} \sim 1.7^{\rm m}$ is required to
produce the observed R-K color of 5.5, requiring the presence of a
significant amount of dust in order to suppress the ultraviolet light
and redden the galaxies.  Although gravitational lensing by a foreground
galaxy or cluster of galaxies could enhance the brightness of the emission,
there is little evidence of such a galaxy or cluster along the line of sight.

The growing number of very red galaxies at high redshift indicates that
there are new populations uncovered only in infrared surveys (e.g.
Graham \& Dey 1996; Malkan, Teplitz, \& McLean 1996; and references
therein).  In a separate paper (Thompson et al. in preparation), we
present statistics that show extremely red objects (R-K$^\prime >
6$) have a sky density of order 500 deg$^{-2}$ for K$^\prime$
$\leq$19.75, so cK39 may, indeed, be part of a larger population.
These results underscore the importance of employing a number of
different techniques for exploring the epoch of early star formation
and demonstrate that a better understanding of these objects is
important to an understanding of galaxy formation in the early universe.

\acknowledgments

We are grateful to T. Herbst and the staff at UKIRT for help with the 
UKIRT observations, obtained through the UKIRT/MPIA collaboration, as 
well as the staff at the W. M. Keck Observatory for their support 
during the Keck runs, and to the staff at Calar Alto for support of 
the original survey.  We profited from discussions with A. Burkert, M. 
Malkan, K. Meisenheimer, C. Steidel,  M. Schmidt, and  S. White during 
the preparation of this paper.  An anonymous referee provided 
thoughtful comments that helped us improve the final manuscript.  SGD 
wishes to acknowledge support from the Bressler Foundation.  This 
research was supported by the Max-Planck-Society.

\clearpage

\begin{deluxetable}{lrrr}
\tablecolumns{4}
\tablewidth{0pc}
\tablecaption{Observed Properties \label{table1}}
\tablehead{ \colhead{Property}     &
            \colhead{B2 0149$+$33} &
            \colhead{cK39E}        &
            \colhead{cK39W}         }
\startdata
R.A. (2000)   & 1:52:34.52
              & 1:52:30.93
              & 1:52:30.83     \nl
Dec  (2000)   & $+$33 50 33.9
              & $+$33 50 55.3
              & $+$33 50 55.3  \nl
Size (arcsec) & \nodata
              & 0.9
              & 0.75           \nl
mK (1$^{\prime\prime}$ apert) & \nodata
              & 20.57$\pm$0.07
              & 20.27$\pm$0.06 \nl
mK (4$^{\prime\prime}$ apert) & K$^\prime=$15.8
              & \multicolumn{2}{c}{18.87$\pm$0.07}  \nl
mR (4$^{\prime\prime}$ apert) & \nodata
              & \multicolumn{2}{c}{24.35$\pm$0.10}  \nl
$\lambda_{\rm line}$ & 2.250$\pm$0.002
                     & \multicolumn{2}{c}{2.248$\pm$0.004} \nl
F$_{\rm line}$ (erg s$^{-1}$ cm$^{-2}$) & 8.4$\times10^{-15}$
   & \multicolumn{2}{c}{6.5$\times10^{-16}$ ($\pm$ 50\%)}  \nl
FWHM ($\mu$m) & 0.0236$\pm$0.0008
              & \multicolumn{2}{c}{0.0165 ($+$0.0075,$-$0.0165)} \nl
FWHM (km s$^{-1}$) & 3150$\pm$100
   & \multicolumn{2}{c}{2200 ($+$1000,$-$2200)} \nl
\enddata
\end{deluxetable}

\clearpage

\begin{deluxetable}{lrr}
\tablecolumns{3}
\tablewidth{0pc}
\tablecaption{Properties of cK39 \label{table2}}
\tablehead{ \colhead{Property}     &
            \multicolumn{2}{c}{Assumed line is:} \\
            \cline{2-3} \\
            \colhead{($h_{50}=1, q_0=0.5$)} &
            \colhead{P$\alpha$}        &
            \colhead{H$\alpha$}        }
\startdata
Redshift & 0.199 $\pm$ 0.002
         & 2.425 $\pm$ 0.007 \nl
Size (kpc): E-W separation & 5.5 & 10 \nl
Survey volume (Mpc$^3$) &  1,555 $h_{50}^{-2}$
                        & 28,500 $h_{50}^{-2}$ \nl
Line luminosity (erg s$^{-1}$)\tablenotemark{a}
   & 1.2$\times10^{41}$ $h_{50}^{-2}$
   & 2.7$\times10^{43}$ $h_{50}^{-2}$ \nl
Eq. Width (\AA) & 360 & 125 \nl
SFR (M$_\odot$ yr$^{-1}$) & 9 $h_{50}^{-2}$
                          & 240 $h_{50}^{-2}$ \nl
A$_{\rm V}$ (mag) & 4 $-$ 8 & 0 $-$ 3 \nl
$\rho$ (Gpc$^{-3}$) & 3$\times10^4$ $-$   3$\times10^6$
                    & 1.6$\times10^3$ $-$ 1.6$\times10^5$ \nl
\tablenotetext{a} {The calibration uncertainties are $\sim$50\%.}
\enddata
\end{deluxetable}

\clearpage

\begin{deluxetable}{lrrrrr}
\footnotesize
\tablecolumns{6}
\tablewidth{0pc}
\tablecaption{Comparison of Object Properties \label{table3}}
\tablehead{ \colhead{Property}   &
            \colhead{cK39}       &
            \colhead{MTM 095355} &
            \colhead{HR 10}      &
            \colhead{MS 1512$-$} &
            \colhead{Hawaii 167}  \\
                                 &
                                 &
            \colhead{$+$545428}  &
                                 &
            \colhead{cB58}       &
                                   }
\startdata
Redshift & 2.43 & 2.5 & 1.44 & 2.72 & 2.36 \nl
$\Sigma$ (objects deg$^{-2}$) & 14 -- 1400
                              & 150 -- 15,000
                              & 1.7 -- 170
                              & 0.11 -- 11
                              & 2.2 -- 220 \nl
K$^\prime$ & 18.5 & K$=$19.5 & K$=$18.4 & 17.9 & 17.2 \nl
Color & R-K$^\prime$ $=$ 5.5 & I-K $=$ 4.2 & I-K $=$ 6.5
      & I-K$^\prime$ $=$ 2.0 & I-K$^\prime$ $=$ 2.8 \nl
Size (kpc)/ $h_{50}^{-2}$ & 16$\pm$3 & $\sim$5 & 5.6 & 23 & $<$4 \nl
L$_{\rm H}\alpha$ (erg s$^{-1}$)/ $h_{50}^{-2}$
   & 2.7$\times10^{43}$ & 1.1$\times10^{43}$ & 1.2$\times10^{43}$
   & 3.2$\times10^{43}$ & 2.4$\times10^{44}$ \nl
Rest Weq (\AA) & 125 & 102 & 600 & 58 & 320 \nl
FWHM (km s$^{-1}$) & $\sim$3000 & $<$3000 & 7000 $\pm$ 3000
   & \nodata & 5000 \nl
SFR (M$_\odot$ yr$^{-1}$)/ $h_{50}^{-2}$ & 240 & 100 & $>$100
   & $>$290 & $>$2000 \nl
\enddata
\end{deluxetable}

\clearpage

\figcaption[Beckwith.newfig1.ps]{The top portion of the figure shows the
            two-dimensional spectra of the quasar's H$\alpha$ line (scaled 
            to show the peak of the emission) and the same spectral region 
            at the position of cK39.  Despite the relatively poor 
            signal-to-noise ratio for cK39, the emission line is apparent 
            in the spectrum.  The bottom panel shows the extracted spectra 
            of cK39 (solid histogram) and the quasar (dotted line) at full 
            resolution.  The quasar spectrum has been scaled down by a 
            factor of 30 to overplot cK39.  The dashed line shows the zero 
            flux level.
            \label{fig1}}

\figcaption[Beckwith.newfig2.ps]{The top part of the figure is a portion of
            the R band image from the Low Resolution Imaging Spectrograph
            (Oke et al. 1995) on Keck, centered on the candidate, cK39.
            The bottom part of the figure is the K band image, obtained
            with the Near-Infrared Camera (Matthews \& Soifer 1994) on
            Keck, for comparison.  Each image is 30$^{\prime\prime}$
            square; north is up, east to the left.  Arrows denote cK39 in
            both images.  
            \label{fig2}}

\clearpage

\begin{figure}
\epsscale{.5}
\plotone{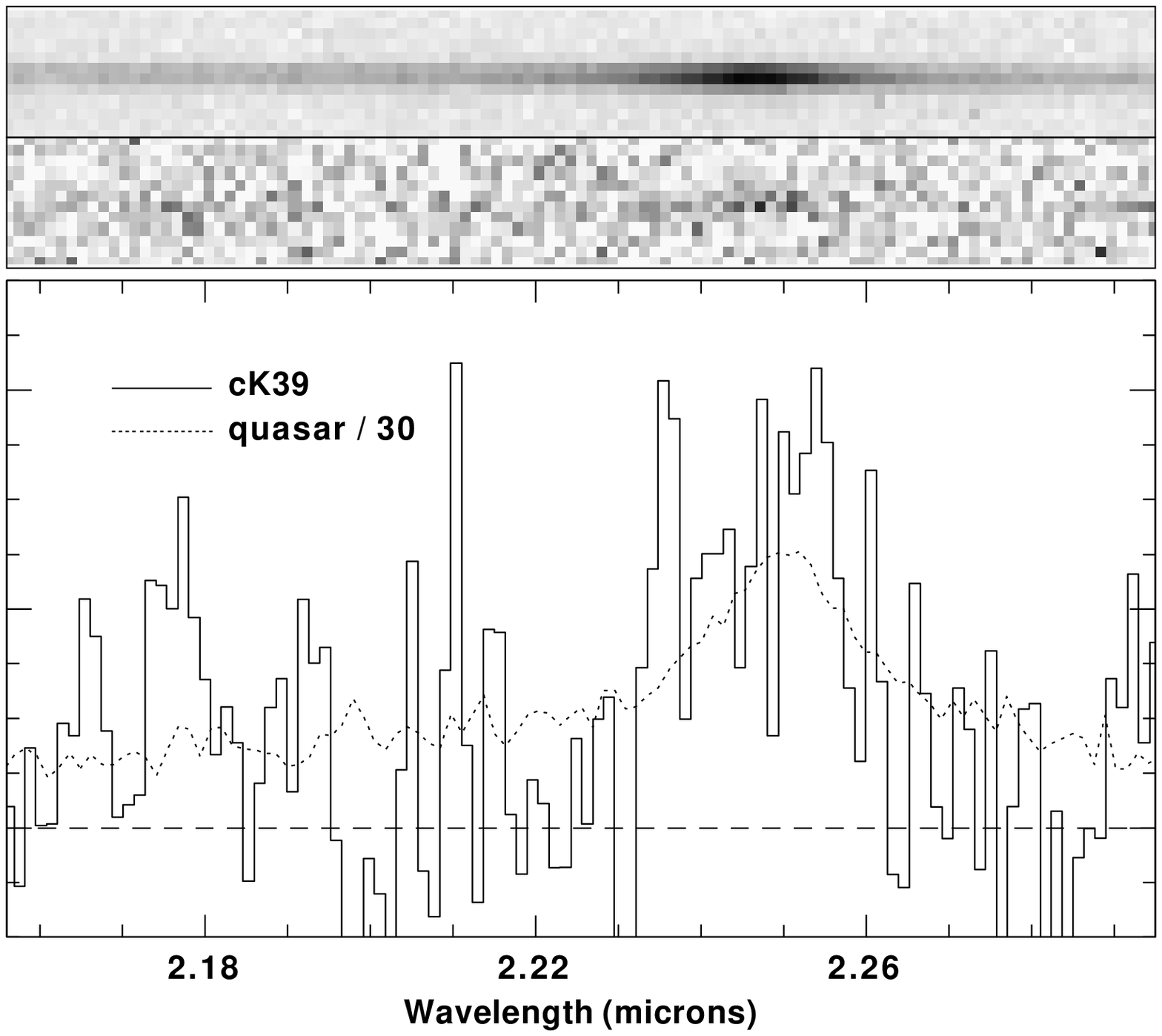}
\epsscale{1.0}
\end{figure}
\clearpage

\begin{figure}
\epsscale{.5}
\plotone{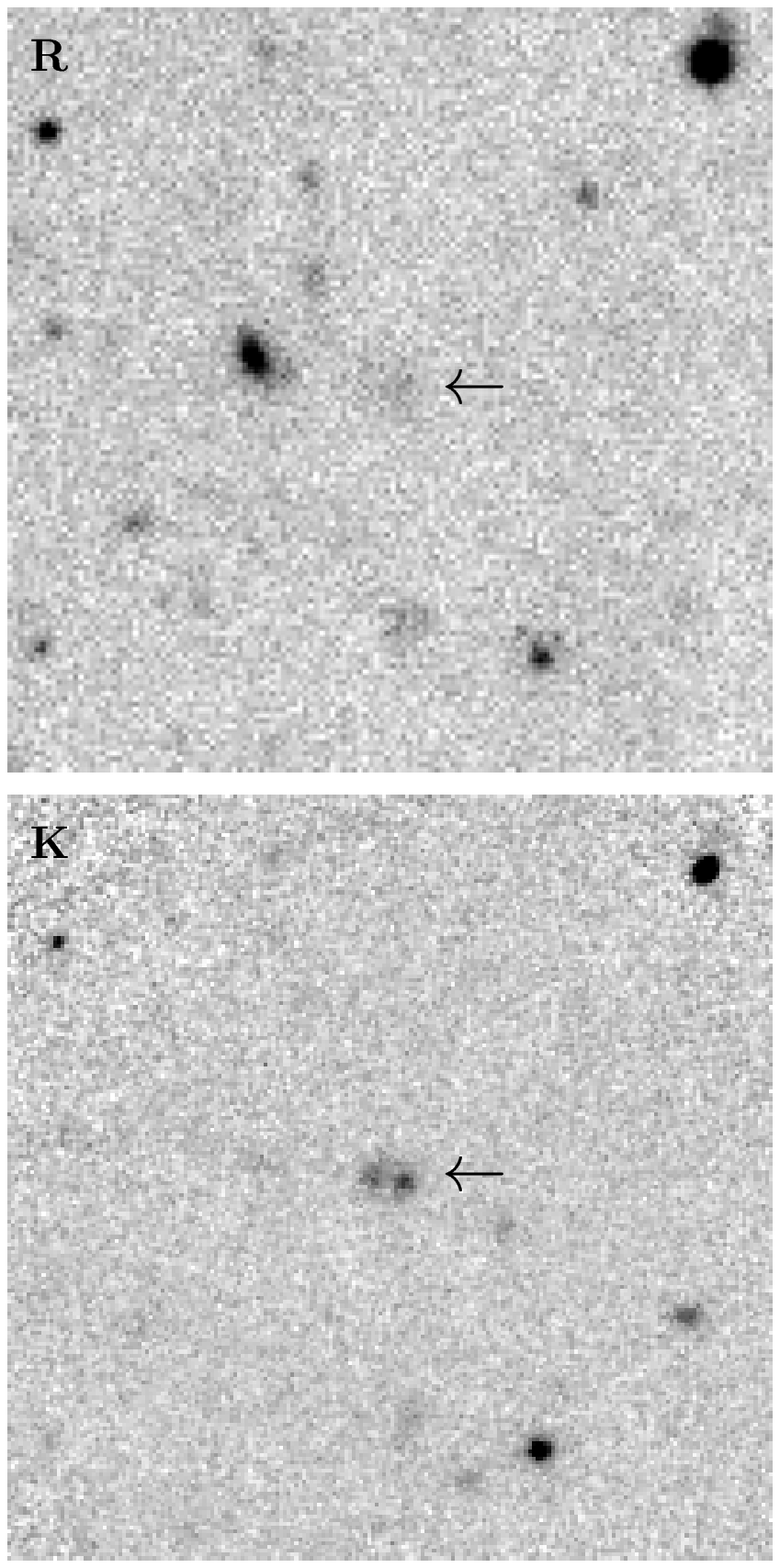}
\epsscale{1.0}
\end{figure}
\clearpage

\end{document}